\documentclass{PoS}

\usepackage{graphicx}
\usepackage{amsmath}
\usepackage{epstopdf}

\def\fm {\mathop{\hbox{fm}}}
\def\MeV {\mathop{\hbox{MeV}}}

\def\Re {\mathop{\hbox{Re}}}

\def\Tr {\mathop{\hbox{Tr}}}
\def\DU  {\mathop{{\mathcal{D}}\hbox{U}}}

\def\dd  {\mbox{d}}
\newcommand\detn[1]{\mbox{det}_{#1}}
\newcommand\tdetn[1]{\widetilde{\mbox{det}}_{#1}}
\newcommand{\beq}{\begin{equation}}
\newcommand{\eeq}{\end{equation}}
\newcommand{\beqa}{\begin{eqnarray}}
\newcommand{\eeqa}{\end{eqnarray}}

\title{Study of QCD critical point using the canonical ensemble method }

\ShortTitle{Study of QCD critical point using the canonical ensemble method }

\author{\speaker{Anyi Li} \footnote{Present address: Department of Physics, Duke University, Durham, NC 27708} \\
        Department of Physics and Astronomy, University of Kentucky, Lexington, KY 40506\\
        E-mail: \email{anyili@phy.duke.edu}}

\abstract{
The existence of the QCD critical point at non-zero baryon density is not
only of great interest for experimental physics but also a challenge for the
theory. Any hint of the existence of the first order phase transition and, particularly,
its critical point will be valuable towards a full understanding  of the QCD phase diagram.
We use lattice simulation based on the canonical ensemble
method to explore the finite baryon density and finite temperature region and look for the QCD critical point.
As a benchmark, we run simulations for the four
degenerate flavor QCD where we
observe a clear signal of the expected first order phase transition.
In the two flavor case, we do not see any signal
for temperatures as low as $0.83\ \rm{T_c}$. Although our real world contains two light quarks and
one heavier quark, three degenerate flavor case shares a lot of similar phase structures as the QCD.
We scan the phase diagram using clover fermions
with $m_\pi \approx 700\mbox{MeV}$ on $6^3\times4$ lattices.
The baryon chemical potential is measured as we increase the baryon number and we see the characteristic ``S-shape''
that signals the first order phase transition. We determine the phase boundaries by Maxwell construction
and report our preliminary results for the location of critical point for the present lattice.

}

\FullConference{The XXVII International Symposium on Lattice Field Theory\\
                 July 26-31, 2009\\
                 Peking University, Beijing, China}

\begin{document}

\section{Introduction}

\begin{figure}[h!]
  \center
  \includegraphics[scale=0.30]{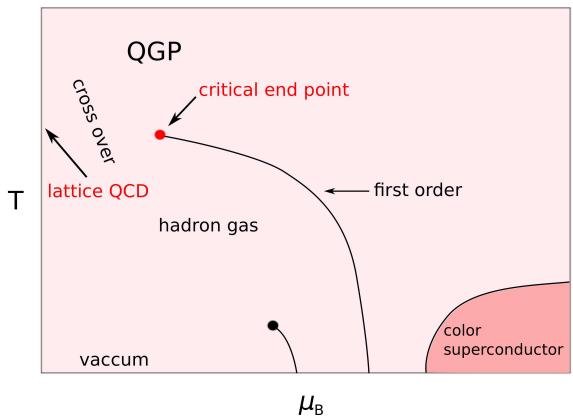}\\
  \caption[A conjectured QCD phase diagram]{Conjectured QCD phase diagram}\label{qcd-phase-diagram}
\end{figure}

Quantum Chromodynamics (QCD) is a fundamental theory which
describes the strong interaction of quarks and gluons---the basic blocks of matter. The
knowledge of QCD at finite temperature and finite baryon chemical potential is essential to
the understanding of a variety of phenomena. Exploring the QCD phase diagram, including identification of different phases and
the determination of the phase transition line is currently one of the most studied topics~\cite{Karsch:2003jg, Muroya:2003qs}. Guided by phenomenology and experiments, a candidate QCD phase diagram was proposed~\cite{Rajagopal:2000wf} -- Fig.~\ref{qcd-phase-diagram} shows a sketch of the conjectured phase diagram. In certain limits, in 
particular for large temperatures $T$ or large baryon-chemical potential $\mu_B$, thermodynamics is dominated by short-distance QCD dynamics and the theory can be studied analytically. However, the most interesting phenomena lie in regions where non-perturbative features of the
theory dominate. The only known systematic approach is the first-principle calculation via lattice QCD  using 
Monte Carlo simulations.

The thermodynamics of strongly interacting matter has been studied extensively in lattice calculations at vanishing baryon
chemical potential. Current
lattice calculations strongly suggest that the transition from the low temperature hadronic phase to
the high temperature phase is a continuous but rapid crossover happening in a narrow temperature interval around  $\rm{T_c} \sim 170 \MeV$~\cite{Aoki:2006we,Cheng:2006qk,Detar:2007as,Karsch:2007dt,Karsch:2008fe,Aoki:2006br,Aoki:2009sc}. On
the other side of the phase diagram~---~large baryon chemical potential but very low temperature, a number of different model approaches~\cite{Alford:1998mk,Brown:1990ev,Asakawa:1989bq,Barducci:1989wi,Barducci:1989eu,Barducci:1993bh,Berges:1998rc,Halasz:1998qr} suggest that the
transition in this region is strongly first order, although this argument is less robust. 
This first order phase transition is expected to become less pronounced as we lower the chemical potential and it
should terminate at a second order phase transition point -- the critical point.

The search for the QCD critical point has attracted considerable theoretical and experimental
attention recently. The possible existence of this point was
introduced sometime ago~\cite{Asakawa:1989bq,Barducci:1989wi}. It is apparent that the location of the
critical point is a key to the understanding of the QCD phase diagram. For experimental search of the critical point, it has been
proposed to use heavy ion collisions at (RHIC) ~\cite{Stephanov:1998dy,Stephanov:1999zu}. The appearance of this point is closely
related to hadronic fluctuations which may be examined by an
event-by-event analysis of experimental results. The upcoming RHIC lower energy scan and future LHC and FAIR (GSI) will focus on
the region $T > 100 \MeV$ and $\mu_B \sim 0 - 600 \MeV$ where the critical point is predicted to exist in theoretical models.
However, in order to extract unambiguous signals for the QCD critical
point, quantitative calculations from first-principle lattice QCD are indispensable.

Remarkable progress in lattice simulations at zero baryon chemical potential has been made
in recent years; however, simulations at non-zero chemical
potential are difficult due to the complex nature of the fermionic determinant at non-zero chemical potential. The
phase fluctuations produce the notorious ``sign problem''.
The majority of current simulations are focusing on the small chemical potential region $\mu_q/T \ll 1$ where the ``sign problem'' appears to be controllable. Most of them are based on the grand canonical ensemble ($T$, $\mu_B$ as parameters).
Since the existence and location of
the critical point is still unknown, we need an algorithm which can be applied beyond the small
chemical potential region. This is one of the motivations of our studies via the canonical ensemble
approach~\cite{Engels:1999tz, Liu:2000dj, Liu:2002qr,Azcoiti:2004ri,deForcrand:2006ec}.

We have proposed an algorithm based on the canonical partition function to alleviate the overlap and 
fluctuation problems~\cite{Liu:2003wy,Alexandru:2005ix}. 
The method we use is computationally expensive since
every update involves the evaluation of the fermionic
determinant; however, finite baryon density simulation based on this method has been shown to be feasible~\cite{Alexandru:2005ix,Alexandru:2007bb}. In this approach, we measure the baryon chemical potential to detect the phase transition.
With the aid of the winding number expansion technique~\cite{Meng:2008hj,Danzer:2009sr,Gattringer:2009wi,Danzer:2008xs}, a program was outlined to
scan the QCD phase diagram in an effort to look for the critical point~\cite{Li:2006qa,Li:2007bj}.

In this study, we present results based on simulations using $6^3\times 4$ lattices with clover fermions.
We have run simulations using two, three and four degenerate flavors of quarks.
As a benchmark, we first study the four flavor case. This system has a first order phase transition that extends all the 
way to zero density and the low density region has been reliable studied using staggered fermions. We map out the phase
transition line using our method and compare it with the results of previous studies; we find very good agreement.
For the two flavor case we present results for three different temperatures, the lowest one being $T=0.83 T_c$. In this
case we do not see any clear signal for a phase transition. While surprising, this is consistent with results from other 
studies~\cite{Ejiri:2008xt}. The most interesting results we report are for the $N_f=3$ case, where the evidence of a
critical point would not only suggest the existence of such a point in the real world but also give an estimation of its location.
We scan the phase diagram and we find signals for a first order phase transition for temperatures below $0.92 T_c$. We 
locate the critical point at the intersection of the boundaries of the coexistence region.

\section{Canonical partition function}

One way to show how to build the canonical ensemble in lattice QCD is to start from the fugacity expansion,
\beq
Z(V,T,\mu) = \sum_{k} Z_C(V, T, k) e^{\mu k/T}, \label{fugacity}
\eeq
where $k$ is the net number of quarks (number of quarks minus number of anti-quarks) and $Z_C$ is the canonical partition function.
Using the fugacity expansion, it is easy to see that we
can write the canonical partition function as a Fourier transform of the grand canonical partition function,
\beq
Z_C(V, T, k) =
\frac{1}{2\pi} \int_0^{2\pi} \mbox{d}\phi \,e^{-i k \phi} Z(V, T,\mu)|_{\mu=i\phi T}.
\eeq
after integrating over the imaginary chemical potential.

For our study, we use the Iwasaki improved gauge action and clover fermions~\cite{Iwasaki:1996ya,AliKhan:2000iz,AliKhan:2001ek}.The imaginary chemical potential is introduced via a $U(1)$ phase added to the time links from the last time slice.

As an illustration, we will consider the case of two degenerate flavors. After integrating out the
fermion part, we get
\beq
Z_C(V, T, k) = \int \DU e^{-S_g(U)} \detn{k}
M^2(U),
\eeq
where
\beq
\detn{k} M^2(U) \equiv \frac{1}{2\pi}\int_0^{2\pi} \dd\phi\,e^{-i k \phi} \det M^2(m,
\mu;U)|_{\mu=i\phi T} ,
\eeq
is the  projected determinant with the fixed net quark number $k$. Using
charge transformation and $\gamma_5$ hermiticity property of the action, we can prove that
\beq
Z_C(V,T,k) = Z_C(V,T,-k).
\eeq
This property allows us to rewrite the partition function as
\beq
Z_C(V,T,k) = \int \DU e^{-S_g(U)} \Re \detn{k} M^2(U).\label{eq:canonical}
\eeq
The integrand is real but not necessarily positive.
To compute observables based on this partition function we rewrite it as:
\beqa
  Z_C(V, T, k) &= &\int \DU e^{-S_g(U)} \detn{k} M^2(U)
  \nonumber \\
  &=&\int \DU e^{-S_g(U)} \mbox{det}M^2(U)W(U)\alpha(U),
\eeqa
where
\beq
W(U) = \frac{\mathop{|\Re}\detn{k}M^2(U)|}{\mbox{det}M^2(U)},
\eeq
and
\beq
\alpha(U) = \frac{\detn{k}M^2(U)}{|\mathop{\Re}\detn{k}M^2(U)|}\label{sign_def}.
\eeq
We separate the phase $\alpha(U)$ and generate an ensemble distributed according to the positive weight 
$|\Re \detn{k}M^2(U)|$;
the phase factor is reintroduced in the observable. To generate this ensemble a
candidate configuration is ``proposed'' by the standard Hybrid Monte Carlo
algorithm~\cite{Gottlieb:1987mq,Duane:1987de,Kennedy:1998cu,Clark:2006fx}, then an accept/reject step is used to correct the
weight. We note that the accept/reject step is based on the determinant ratio which
alleviates the fluctuation problem~\cite{Joo:2001bz,Liu:2002qr,Liu:2003wy} and
enhances the acceptance rate~\cite{Alexandru:2005ix}.

\section{Winding number expansion}

\begin{figure}[h]
\centering
\includegraphics[scale=0.55]{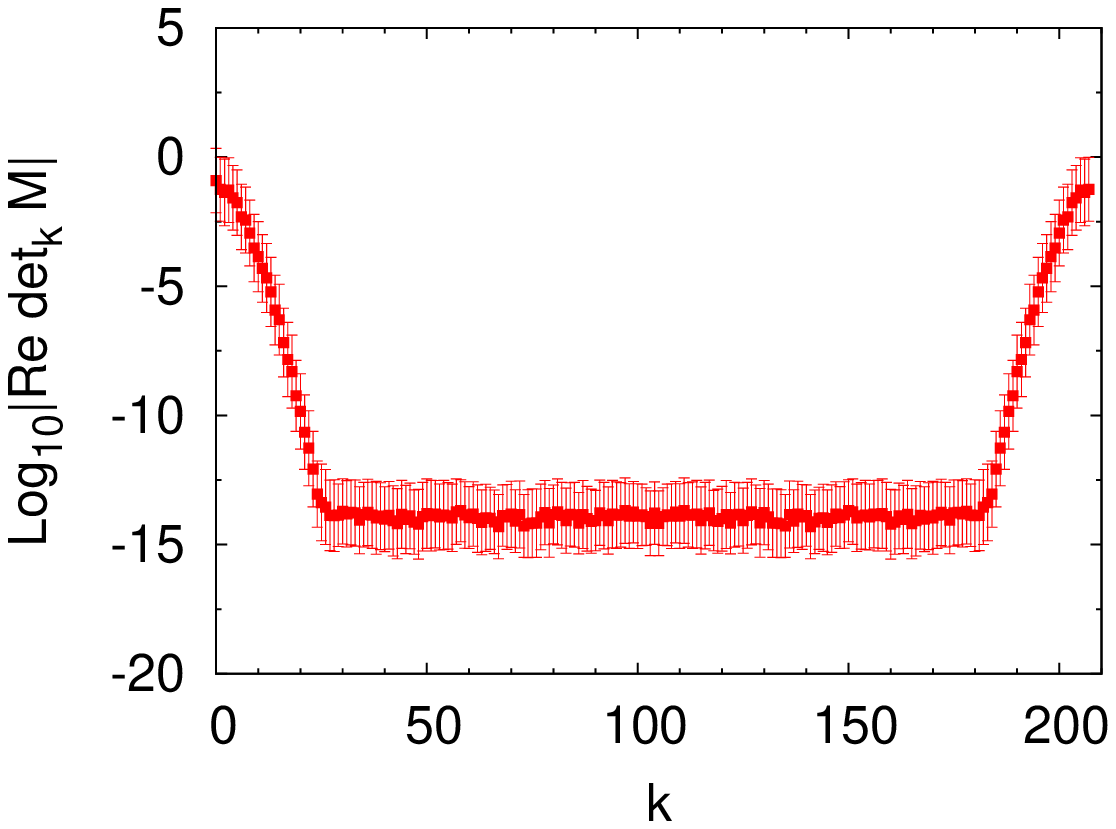}
\includegraphics[scale=0.55]{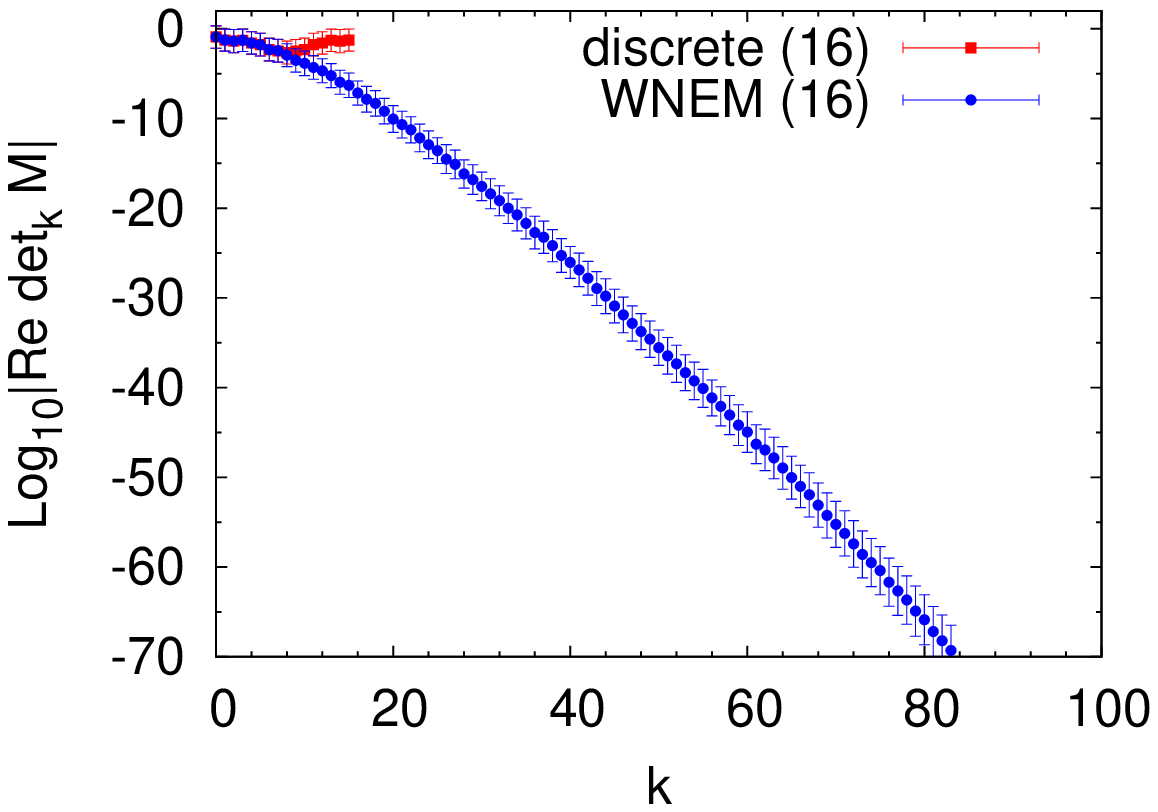}
\caption{Numerical instability of discrete Fourier transform
with 208 points (Left). Comparison from winding number expansion method and the discrete Fourier transform with $N=16$ evaluations (Right). Error bars are the standard deviations.
\label{fig:discrete}}
\end{figure}

Most of the simulation time is spent on the accept/reject step, specifically on computing
determinant of the fermion matrix.
On the $6^3\times4$ lattice,
the matrix has  $10368 \times 10368$ entries. Although it's very sparse, exact determinant calculation
is numerically demanding even on this small lattice. 
Moreover, the determinant needs to be evaluated many times at every accept/reject step since we need
to compute its Fourier transform; our original approach was to use an
approximation where the continuous Fourier transformation is replaced by a discrete one, i.e.:
\begin{equation}
\detn{k} M^2(U) \approx \frac{1}{N} \sum_{j=0}^{N-1} e^{-i k
\phi_j} \det M(U_{\phi_j})^2,~~~~~\phi_j=\frac{2\pi
j}{N}.\label{eq:fourier}
\end{equation}
It was shown that the errors introduced by this approximation are small~\cite{Li:2007bj} for small quark numbers. However, there are two problems
with this approach: (i) the computation time increases linearly with the net quark number,
(ii) although evaluating the determinant $N$ times should theoretically allow us to compute projection numbers as large as
$k=N/2$, numerically we found that for large quark numbers, the Fourier components become too
small to be evaluated with enough precision, even
using double precision floating point numbers. It has been shown~\cite{Meng:2008hj} that the
the results of the projected determinant
for $k$ larger than $20$ would differ significantly for different choice of $N$, which signals a numerical instability.
To see this problem, we use $N=208$
to evaluate the fermion determinant and calculate Fourier projection using discrete Fourier
transform. One would expect that  $|\mathop{\Re}\detn{k}M2(U)|$ should decrease exponentially.
We see that this is indeed
the case for discrete Fourier transform at small quark number, as shown in
Fig.~\ref{fig:discrete} (left panel). However, as the quark number gets close to 30, the
projected determinant calculated using the discrete Fourier transform
approximation flattens out and stops falling
below the double precision limit of $10^{-15}$.
This is the onset of the numerical instability.

This happens because $e^{-i k \phi_j}$ oscillates rapidly at larger $k$, leading to large cancellation in the sum 
over $e^{-i k \phi_j} \det M(U_{\phi_j})^2$.
The accumulation of round-off errors makes it impossible to evaluate the projected determinant.
This numerical challenge led us to develop a new method which
should be free of the above numerical problem for quark numbers relevant to the phase transition 
region in this study~\cite{Meng:2008hj}.
The idea of our new method is to first consider the Fourier transform of $\log \det M(U,\phi)$ instead of
the direct Fourier transform of $\det M(U,\phi)$. Using an
approximation based on the first few components of $\log \det M(U,\phi)$, we can then analytically
compute the projected determinant. The efficacy of the
method can be traced to the fact that the Fourier components of $\log \det M(U,\phi)$ are the number of terms in the expansion which characterizes the number of quark loops wrapping around the time boundary. They are exponentially smaller with increasing winding numbers. This is why we can
approximate the exponent of the determinant very accurately with a few terms which, in turn, allows us to evaluate the Fourier components of the determinant precisely.

To see how this works, we look at the hopping expansion of the $\log\det M(U,\phi)$. We start by
writing the determinant in terms of the trace log of the quark matrix
\beq
\label{trlog}
\det M(U,\phi)=\exp(\Tr\log M(U,\phi)).
\eeq

It is well known that $\Tr\log M$ corresponds to a sum of quark loops. We separate all these loops in classes depending on how many times 
they wrap around
the lattice in the temporal direction. We have then
\beqa
\label{quark loop}
 &\Tr \log M(U,\phi)&=\sum_{loops}L(U,\phi)\\{\nonumber}
 &&= A_{0}(U)+
 [\sum_{n}e^{in\phi}W_{n}(U)+e^{-in\phi}W_{n}^{\dag}(U)],
\eeqa
where $n$ is the  winding number of quark
loops wrapping around the time direction and $W_{n}$ is
the weight associated with all these loops with winding number $n$. $A_0(U)$ is the contribution with zero winding number.
Eq.~(\ref{quark loop}) can be re-written as
\beqa
\label{wne expression}
 &\Tr \log M(U,\phi)&= A_{0}(U)+
 [\sum_{n}e^{in\phi}W_{n}(U)+e^{-in\phi}W_{n}^{\dag}(U)]\\{\nonumber}
 &&=A_{0}(U)+\sum_{n} A_{n}
 \cos(n\phi+\delta_{n}),
\eeqa
where $A_n \equiv 2|W_n|$ and $\delta_n \equiv \arg (W_n)$ are independent of $\phi$.
Using Eq.~(\ref{trlog}) and Eq.~(\ref{wne expression}) we get
\beq
\label{logdet wne} \det
M(U,\phi)=e^{A_{0}+A_{1}\cos(\phi+\delta_{1})+A_{2}\cos(2\phi+\delta_{2})+.....}.
\eeq

The Fourier transform of the first order in the expansion can now be computed analytically; we have
\beq
\int_{0}^{2\pi}\frac{d\phi}{2\pi}e^{-ik\phi}e^{A_0+A_{1}\cos(\phi+\delta_{1})}=e^{A_0+ik\delta_{1}}{
I}_k(A_1),
\eeq where ${I}_k$ is the Bessel function of the first kind with rank $k$.

For higher orders in the winding number expansion, we compute the
Fourier transform based on the truncated Taylor expansion:
\begin{eqnarray}
&&\int_{0}^{2\pi}\frac{d\phi}{2\pi}e^{-ik\phi}e^{A_0+A_{1}\cos(\phi+\delta_{1})}
e^{\sum_{k=2}^{\infty}A_{k}\cos(k\phi+\delta_{k})}
{\nonumber}\\
&=&\int_{0}^{2\pi}\frac{d\phi}{2\pi}e^{-ik\phi}e^{A_0+A_{1}\cos(\phi+\delta_{1})}
\prod_{k=2}^{\infty} \sum_{n_k=0}^{\infty}\frac{A_k^{n_k}}{n_k!} \cos(k\phi+\delta_k)^{n_k} \nonumber\\
&=&c_{00}{\mathop{\hbox{I}}}{_{k}}(A_{1})+c_{+01}{\mathop{\hbox{I}}}{_{k+1}}(A_{1})+c_{-01}{\mathop{\hbox{I}}}{_{k-1}}(A_{1})+c_{+02}{\mathop{\hbox{I}}}{_
{k+2}}(A_{1})+...
\label{final wne}
\end{eqnarray}
The projected determinant is written in terms of the linear combination of Bessel functions,
the coefficients $c$ can be easily computed
analytically. Using Eq.~\eqref{final wne} and the recursion relation for the Bessel function,
${\mathop{\hbox{I}}}{_{k-1}}(A)=\frac{2k}{A}{\mathop{\hbox{I}}}{_{k}}(A)+{\mathop{\hbox{I}}}{_{k+1}}(A)$, 
the winding number expansion method (WNEM) can be extended to higher orders.

As we mentioned before, the efficacy of the method rests on
the fact that we can get a very good approximation of the exponent using only a few terms in the Fourier 
expansion. However, the evaluation in Eq.~\eqref{final wne} is based on
a mixture of analytical integration
and a truncated Taylor expansion. Since the error of the truncated Taylor expansion is not well
controlled, to produce our final results, we decided to employ a method which can
evaluate Eq.~\eqref{final wne} more precisely
and thus correct the necessary errors introduced by the Taylor approximation. The methodology can be viewed as a form of reweighting and the approximation level is then tuned to produce errors that are smaller than the statistical errors
of the final result. The strategy is the following:
\begin{itemize}
\item
During the procedure of generating ensembles, we compute the determinant for 16 phases
and we keep only 6 winding loops in the $\Tr \log M$ expansion which has been shown to be precisely enough for the approximation of $\Tr \log M$~\cite{Meng:2008hj}. The evaluation of the projected determinant is done by truncating the second and third winding loops terms to their tenth 
order, and first order for the remaining terms. We denote the
approximated determinant using truncated
Taylor series in Eq.~\eqref{final wne} by $\tdetn{k}M$.
\item
Once the ensembles are generated, we evaluate the determinant of each configuration for $N_0$ phases which
admits projection up to support as large us $N_0/2$ winding loops reliably. Instead of evaluating the integral
Eq~\eqref{final wne} by the truncated Taylor series, we evaluate it numerically through
a multi-precision library (GMP library) by setting precision as 512 digits after decimal point.
In order to choose $N_0$, we include the higher order of winding loops until the projected determinant converges, the minimal number of winding loops included will be equal to $N_0/2$.
\item
Since the ensemble generated in the Monte Carlo process involves an approximation of the projected determinant and is not the target ensemble, we correct for this in the observable with the following reweighting
\begin{equation}
\langle O \rangle = \frac{\langle O \alpha_r \rangle_0}{\langle \alpha_r \rangle_0}
\end{equation}
where $\langle \rangle_0$ denotes the average over the ensemble generated with the approximated determinant
and
\begin{equation}  \label{sign_rew}
  \alpha_r = \frac{\detn{k} M}{\tdetn{k} M}
\end{equation}
is the reweighting phase where $\detn{k} M$ is from the exact Fourier transform without Taylor expansion. If the approximation due to Taylor expansion is very good then $\langle \alpha_r \rangle_0$ will be very close to 1; if the difference between the truncated determinant and the exact one is much smaller than the error bars on $\langle O \rangle$ then the reweighting step is not really
needed and we can consider the approximation exact.
On the other hand, if the average is significantly different from 1 we
need to do reweighting. The worst case is when the approximation is so bad as to introduce a sign
problem ($\langle \alpha_r \rangle_0$  overlaps with zero within the error bars). In that case the approximation
is invalid since the generated ensemble
cannot be used to compute any observable.
\end{itemize}
The detailed discussion of this methodology will be presented somewhere~\cite{Ali:2010}. 
By employing the above strategy, even if the approximation is not perfect, it can be corrected by a
reweighting after the ensemble is
generated. The only possible issue is to make sure that the reweighting phase doesn't introduce
too much noise in the observable; if that is the case, the approximation needs to be improved
and a new ensemble needs to be generated.

Before we conclude this section, we would like to compare the merits of WNEM with the those of discrete Fourier
transform. We compute the values of the projected determinant determined using the discrete
Fourier transform with only $N=16$ in Eq.~\eqref{eq:fourier}.
In Fig.~\ref{fig:discrete} (right panel), these results are labeled discrete (16).
We see that this approximation is only valid up to $k=6$.
This is to be compared to WNEM (16) which takes
the same computational time and yet does not suffer from this
problem and, as expected, the evaluated determinant continues to decrease as the quark number is increased.
It is this projected determinant evaluated from WNEM
that allows us to scan a wide range of densities on the QCD phase diagram.

\section{Results}

\subsection{Baryon chemical potential}

Before starting the discussion on the QCD phase diagram, we will first
present the ``tool'' we used to determine the phase transition in the
canonical ensemble: the baryon chemical potential. We measure the chemical potential
and plot it as a function of the net quark number $k$. 
Due to the contribution of the surface tension to the free energy at finite volume,
the chemical potential in the mixed phase region has an ``S-shape'' structure.
We scan the phase diagram looking for this S-shape a structure to signal
a first order phase transition.

In the canonical ensemble, the baryon chemical potential is measured by taking the difference of the free energy after adding one baryon, i.e.
\begin{equation}
\left<\mu\right>_{n_B} = \frac{F(n_B+1)-F(n_B)}{(n_B+1)-n_B} = -\frac{1}{\beta}\ln
\frac{Z_C(3n_B+3)}{Z_C(3n_B)} = -\frac{1}{\beta}\ln \frac{\left<\gamma(U)\right>_o}{\left< \alpha(U)\right>_o}
\label{baryon chemical potential}
\end{equation}
where
\begin{eqnarray}
\alpha(U) &=& \frac{\Re\detn{3n_B} M^{n_f}(U)}{\left|\Re \detn{3n_B}
M^{n_f}(U)\right|}, \quad {\mbox{and}} \\
\gamma(U)&=&  \frac{{\Re\,}\detn{3n_B+3} M^{n_f}(U)} {\left|{
\Re\,}\detn{3n_B} M^{n_f}(U)\right|}.
\end{eqnarray}
are phase factors measured in the ensemble with $n_B$ baryon number where $\left<\right>_o$ in
Eq.~\eqref{baryon chemical potential} stands for the average over the ensemble generated with the measure 
$\left|\Re \detn{3n_B}M^{n_f}(U)\right|$.

\subsection{Phase diagram for $N_f=4$ and $N_f=2$}

We turn now our attention to the QCD phase diagram.
Although it has been speculated for quite a some
time that the QCD phase diagram may look like Fig~\ref{qcd-phase-diagram}, unfortunately, 
there is little solid evidence to to support this scenario. The finite temperature studies at zero
baryon density have shown that the transition from hadronic matter to quark gluon plasma is 
a rapid cross-over. Lattice QCD evidence regarding the existence and location of the second
order transition point will help tremendously in sharpening our understanding of the QCD phase
diagram.

\begin{figure}
\centering
\includegraphics[scale=0.4]{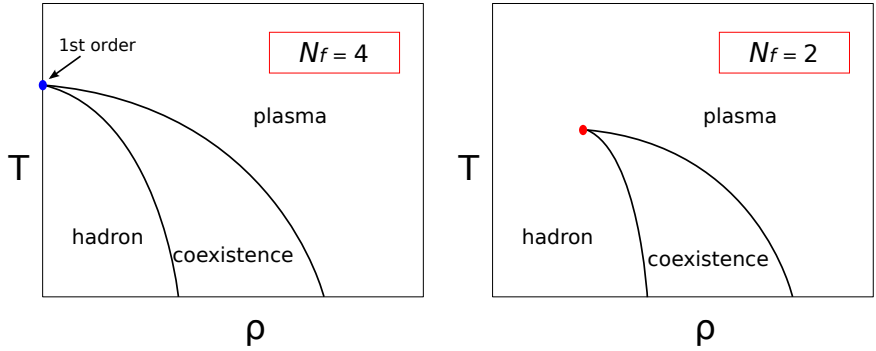}
\caption{Schematic phase diagram of four and two flavors in canonical ensemble}\label{fig:nf2nf4_phase}
\end{figure}

Before we present our results, we would like to point out the difference between the phase diagram in
the grand canonical ensemble and the one in the canonical ensemble. We plot the canonical partition
function phase diagram in Fig.~\ref{fig:nf2nf4_phase}. In terms of $T$ and $\rho$, the first order phase
transition line becomes a phase coexistence region which has two boundaries that separates
it from the pure phases. As we approach the critical point, the two boundaries will eventually
meet at that point. We determine the boundaries of the coexistence phase at different
temperatures and locate the critical point by extrapolation.

As a first attempt, we carry out simulations with four and two degenerated quark flavors.
The reason we start with two and four flavors is the following:
first of all, it is easier to simulate an even number of flavors with HMC~\cite{Duane:1987de}.
Secondly, in the four-flavor case, the first order phase transition line extends all the way to zero density.
This provides a testing ground for the algorithm since there are reliable results mapping the transition line in the
small chemical potential region. Thirdly, the two-flavor case is expected to have a phase diagram similar to that of real QCD. 
The expected phase diagrams for two and four flavor cases are shown in Fig.~\ref{fig:nf2nf4_phase}.
For this reason, we will use the four flavor simulation as a benchmark to demonstrate the
methodology of determining the phase boundaries and locating the critical point before tackling the more realistic case. The expected phase diagrams for two and four flavor cases are shown in Fig.~\ref{fig:nf2nf4_phase}.

Given the first order transition at zero chemical potential for $N_f=4$, it is natural to ask whether
this first order is preserved when we switch on the chemical potential. As we mentioned
before, the first order phase transition is reflected in an ``S-shape'' structure in the $\mu_B$ vs.
$\rho$ plot (or $\mu_B$ vs. $n_B$ plot as we will present). To proceed in this way, we decide to scan the phase diagram by fixing the temperature below 
$\rm{T}_c$  while varying $n_B$.

We scan the phase diagram by fixing the temperature below $\rm{T}_c$  while varying $n_B$. 
Once we cross into the coexistence region, in finite volume, the non-zero contribution from the surface
tension causes the appearance of a ``double-well'' effective potential~\cite{Ejiri:2008xt} which
leads to an S-shaped behavior in the chemical potential plot.
In the thermodynamic limit, the surface tension contribution goes away since it is a surface
term and the free energy scales with the volume;
the chemical potential will then stay constant in the mixed-phase region.

\begin{figure}[hbt!]
\centering
      \includegraphics[scale=0.39]{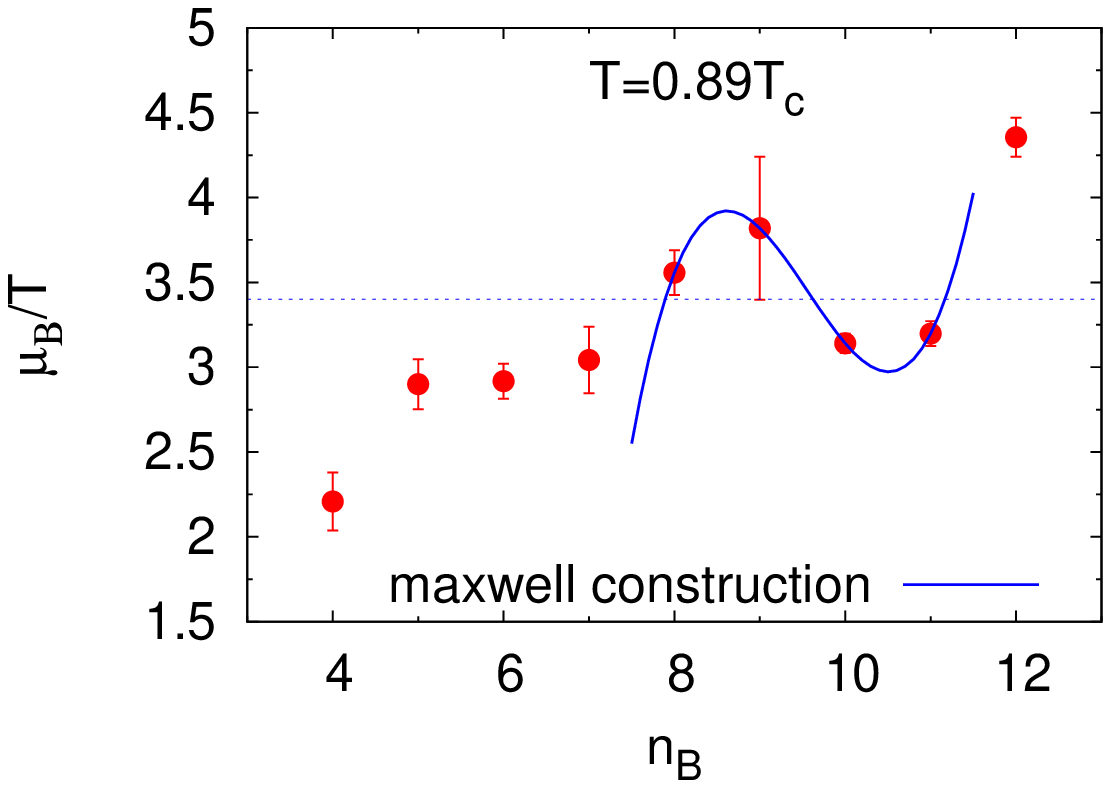}
      \includegraphics[scale=0.39]{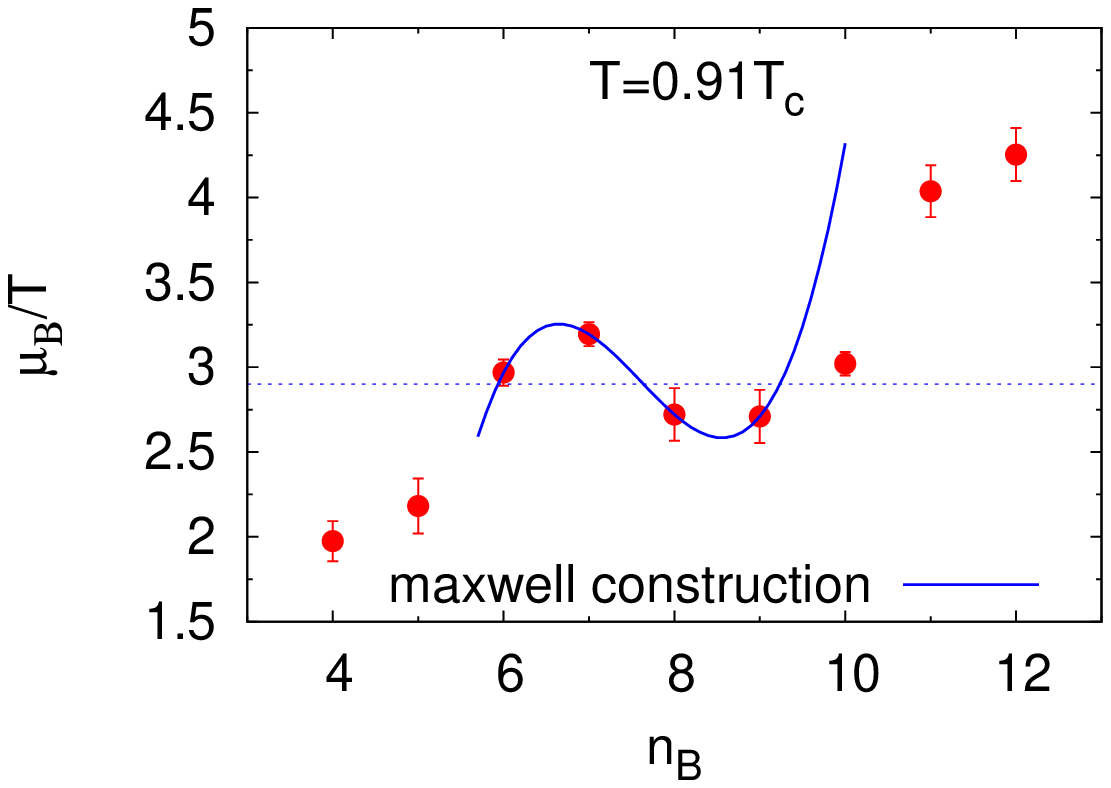}
      \includegraphics[scale=0.39]{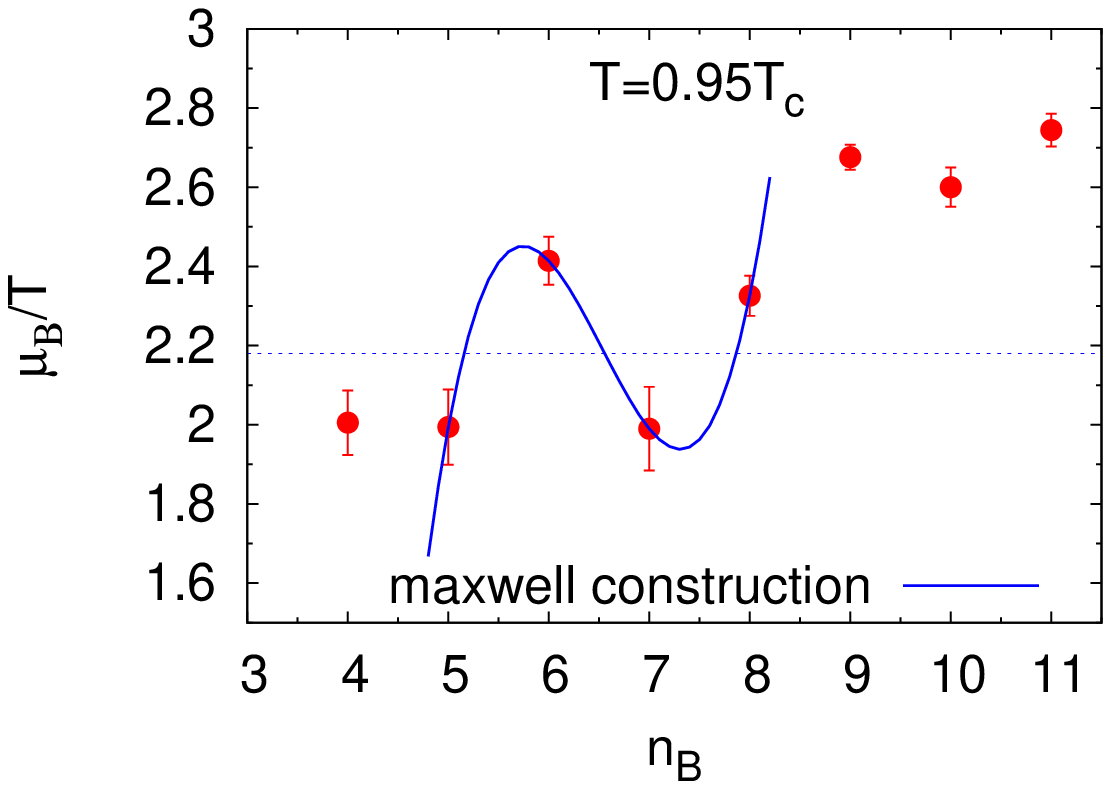}
\caption{$N_f=4$ phase scan along with Maxwell constructions}\label{fig:scaning_nf4}
\end{figure}

We present our results for $N_f=4$ in Fig.~\ref{fig:scaning_nf4} for three different temperature below $\rm{T}_c$. 
To identify the boundaries of the coexistence region and the critical value for the baryon
chemical potential, we rely on the ``Maxwell construction''~\cite{deForcrand:2006ec}: we
select several points in the S-shape region and fit them with a third-order polynomial.
We determine the critical chemical potential $\mu_c$ and the boundary points $\rho_1$ and $\rho_2$
by requiring the two areas between the fitted curve and a constant horizontal line which intersects 
the curve at $\rho_1$ and $\rho_2$ to be equal. For all reasonable fits, we find the values of the 
boundary points, $\rho_1$ and $\rho_2$, and the value of the critical chemical potential ($\mu_c$) to be
fairly insensitive to our choice of the fit function or fit region. The simple third order polynomial fit is sufficient.

\begin{figure}[hbt!]
\centering
\includegraphics[scale=0.7]{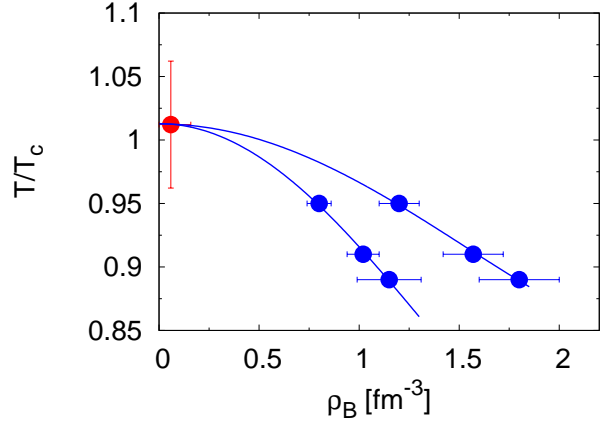}
\caption{Phase boundaries in the temperature vs.\ density plot for
$N_f=4$.}\label{fig:bound_nf4}
\end{figure}

We perform the Maxwell constructions for the three temperatures we studied: $0.89$, $0.91$ and $0.95 \rm{T_c}$;
the results are presented in Fig.~\ref{fig:scaning_nf4}.
Having determined the $\rho_1$ and $\rho_2$ for three different temperatures, we
plot boundaries of the coexistence region by a simple extrapolation.
Although there is no "critical point"
for the $N_f=4$ case, it is expected that
the two coexistence phase boundaries should cross at zero chemical potential and $\rm{T=T_c}$.
To determine the crossing point, we fit the boundary lines using a simple even polynomial in baryon density
\beq
\frac{\rm{T_c(\rho)}}{\rm{T_c(0)}} = 1 - a(N_f,m_q)(V\rho)^2+O\left((V\rho)^4\right)
\label{eq:expansion}
\eeq
to do the extrapolation. The phase boundaries and their extrapolation are plotted in Fig.~\ref{fig:bound_nf4}.
We find the intersection point at $\rm{T_c(\rho)}/\rm{T_c(0)} = 1.01(5)$ and
$\rho=0.05(10)\fm^{-3}$ which is consistent with our expectation: $\mu=0$ and $\rm{T=T_c}$.

We would like to point out that the critical temperature was determined using a different set of simulations.
We run simulations at zero baryon density and we monitored the
Polyakov loop susceptibility as we increased the temperature. These simulations were performed
using the standard HMC algorithm and we used two lattice sizes: $6^3\times 4$ and
$10^3\times 4$. The critical temperature was determined using the peak of the Polyakov loop susceptibility
and we also checked that the susceptibility peak scales with the volume as expected. The fact that the intersection
point is close to the critical temperature determined from a set of simulations using a different methodology
constitutes a convincing cross-check of the correctness of our method.

\begin{figure}[h!]
\includegraphics[scale=0.65]{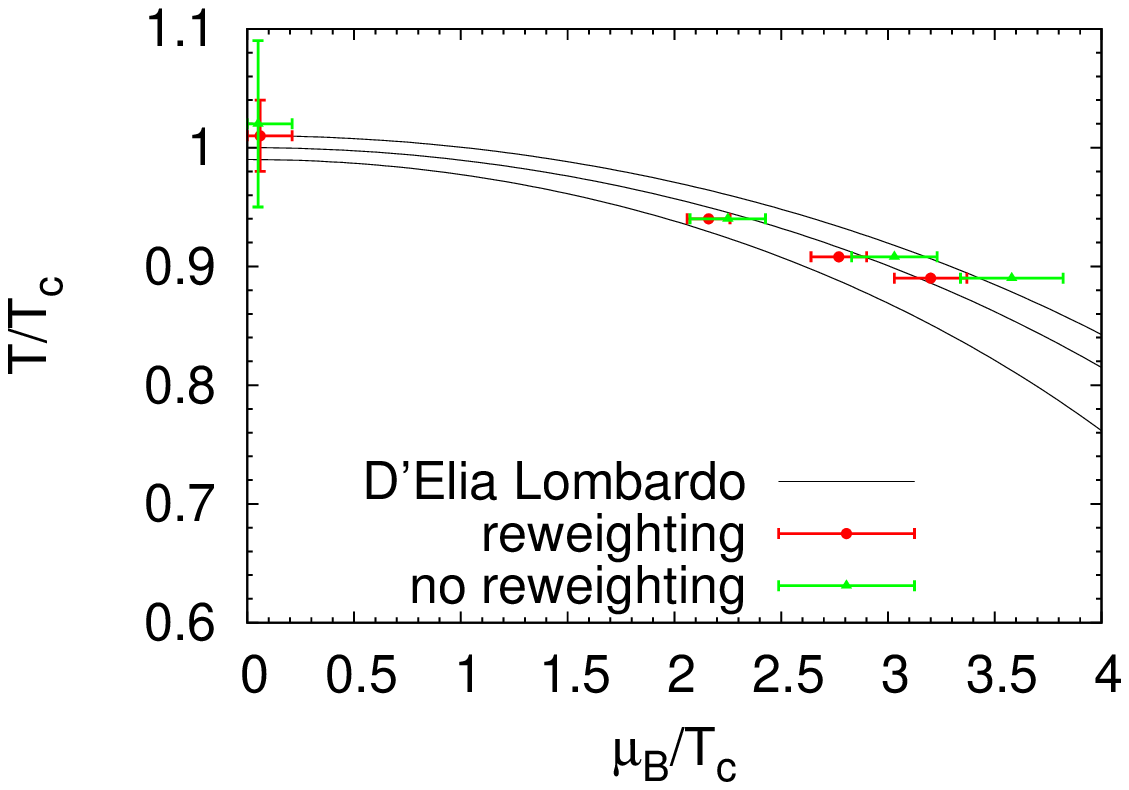} 
\includegraphics[scale=0.27]{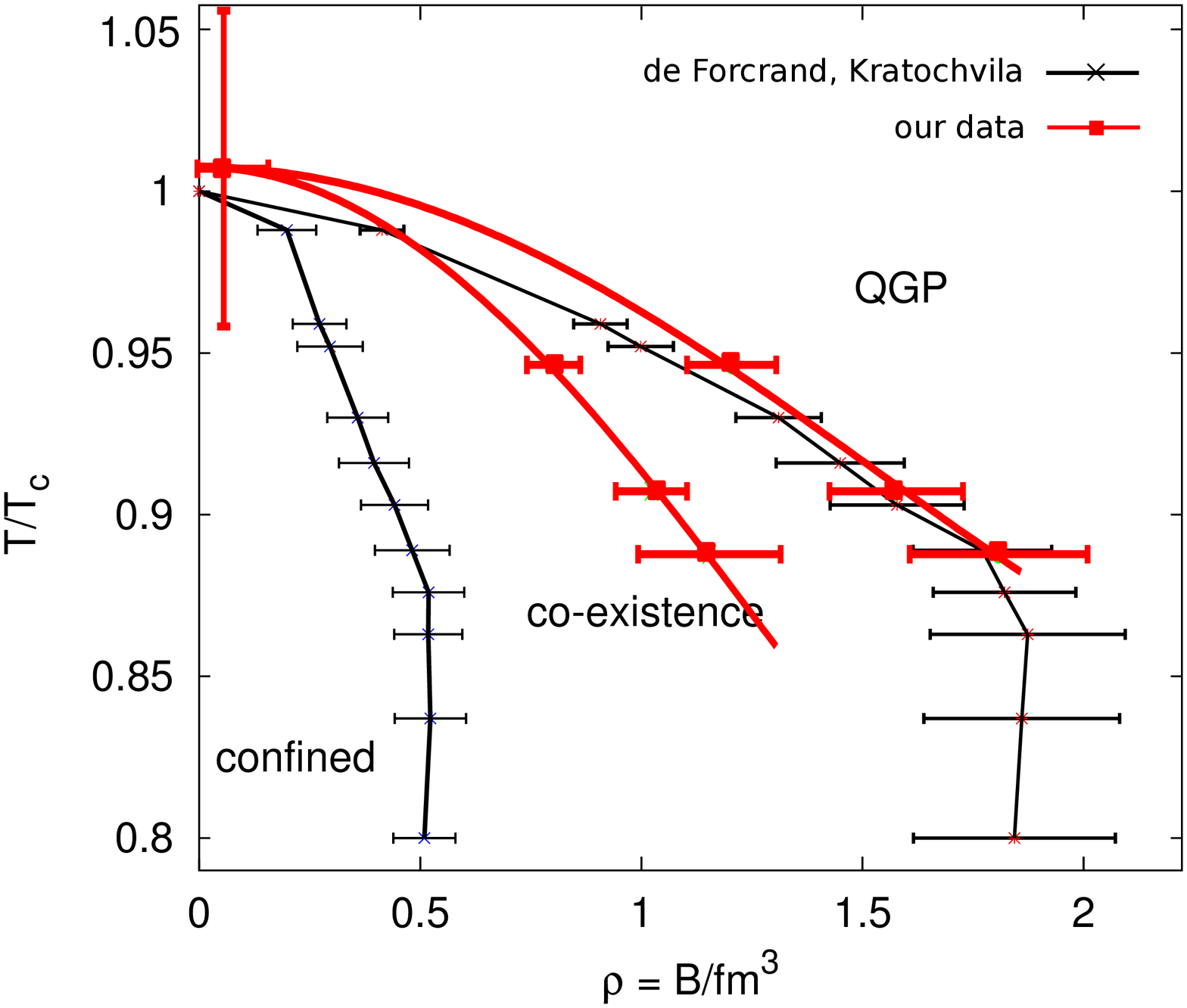}
\caption{Phase transition line for $N_f=4$ in the $T$, $\mu$ plane.}\label{fig:nf4_compare}
\end{figure}

We compare our results to the study of analytic continuation from the imaginary chemical potential with stagger fermions~\cite{D'Elia:2002gd}. The results are presented in Fig.~\ref{fig:nf4_compare} (left panel) -- we see that the agreement is quite good in spite of the
fact that the quark mass used in the staggered fermion study, $m_\pi \approx 300-400\MeV$, is significantly 
different from the one we use. This is not surprising since the phase transition curve for $N_f=4$ is fairly independent of the quark mass~\cite{Philipsen:2008gf}. As a check that the reweighting does not change drastically the observable results,
we also plot the results with no reweighting which should contain systematic errors from Taylor expansion. Nevertheless, we see that the errors of
the chemical potential from non-reweighted results overlap with those from the reweighted ones and the extrapolated crossing point of the two
boundaries also agree with $T_c$ at $\mu =0$ albeit with larger errors.
\begin{figure}[hbt!]
\centering
      \includegraphics[scale=0.55]{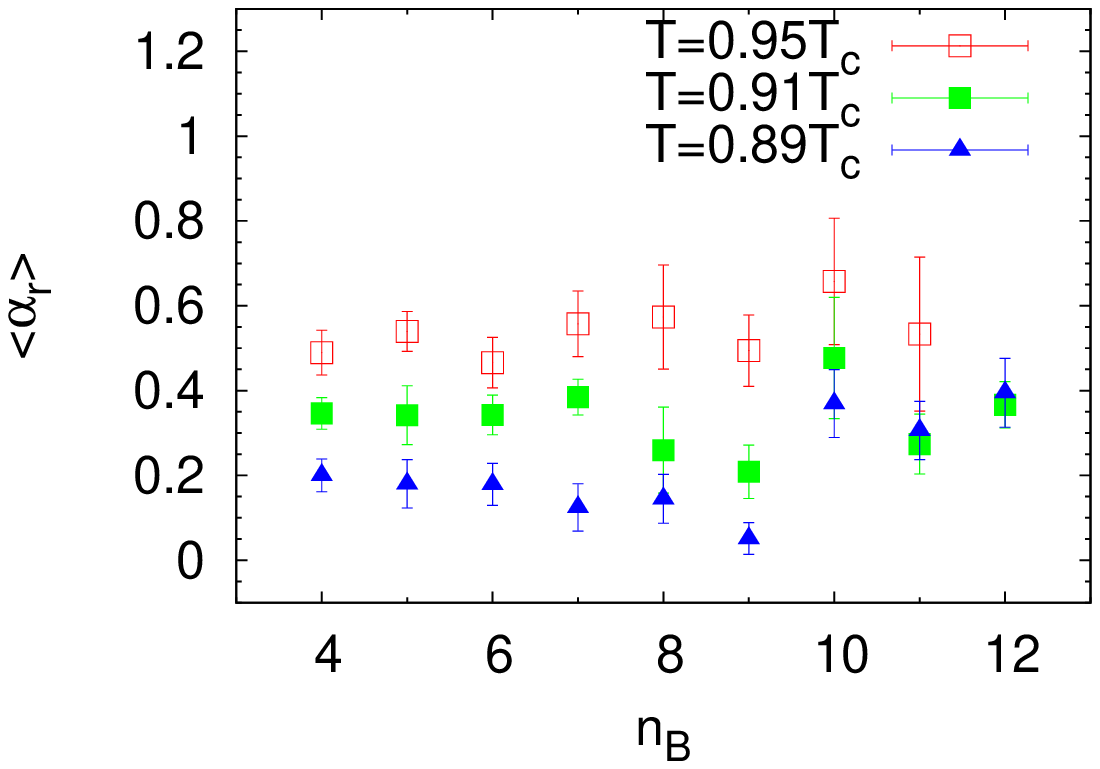}
      \includegraphics[scale=0.55]{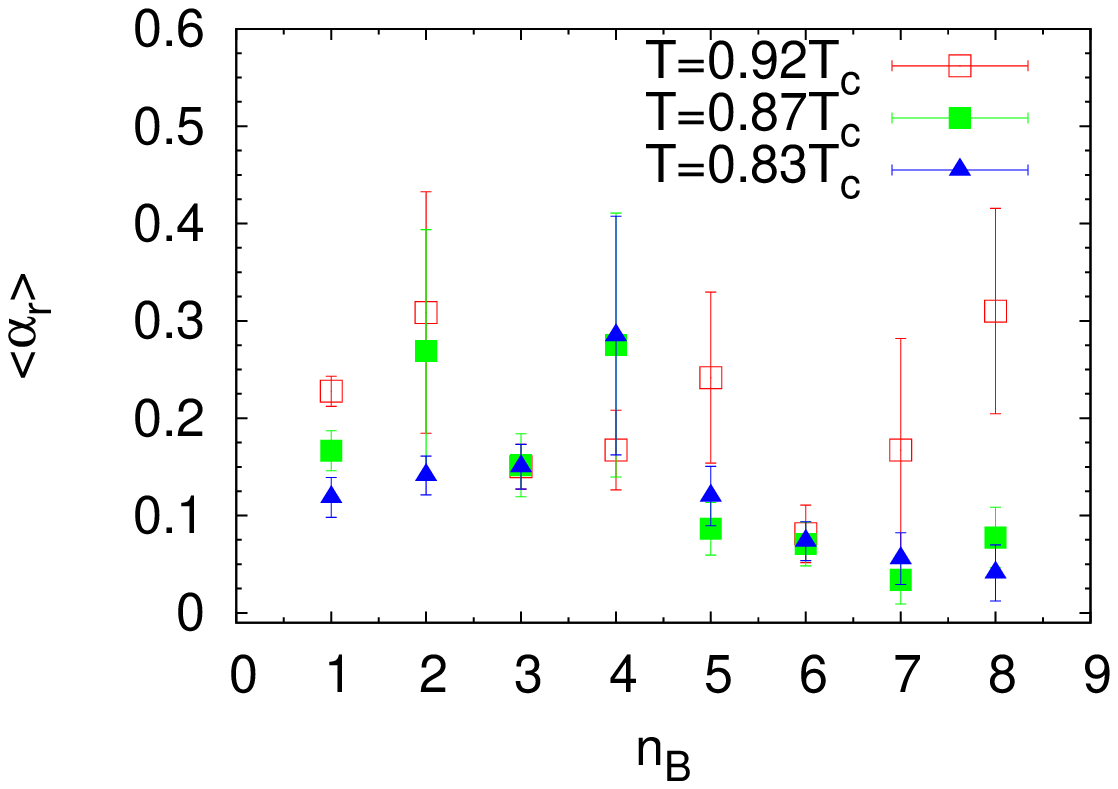}
\caption{Left: Average sign of $N_f=4$ in terms of $n_B$. Right: Average sign of $N_f=2$ in terms of $n_B$}\label{fig:sign_nf24}
\end{figure}
We also compare our results to ones from a canonical ensemble study that uses staggered fermions~\cite{deForcrand:2006ec} (Fig.~\ref{fig:nf4_compare} (right panel)).
We find that the results are consistent with each other; however, our coexistence region is narrower. 
The discrepancy could come from the difference in the quark mass 
($m_\pi \approx 800 \MeV$ for our study compared to $m_\pi\approx 300 \MeV$ for the staggered study), 
and/or the fact that we use a different fermion formulation.

As a sanity check, we examine the seriousness of the potential sign problem. The average sign in Eq.(\ref{sign_rew}) is plotted in Fig.~\ref{fig:sign_nf24} (left panel).
We see that they range mostly from 0.6 to 0.2. Except for the point at $\rm{T}=0.89 \rm{T_c}$ and $n_B=9$ which has a 1.5 sigma away from zero, the others
all have more than 3 sigmas above zero. In view of the fact that our results agree with those from the imaginary chemical potential study and
the extrapolated boundaries meet at the expected $T_c$ at $\mu =0$, we believe that the sign fluctuation in the reweighting is not a problem. 

\begin{figure}[hbt!]
\centering
\includegraphics[scale=0.39]{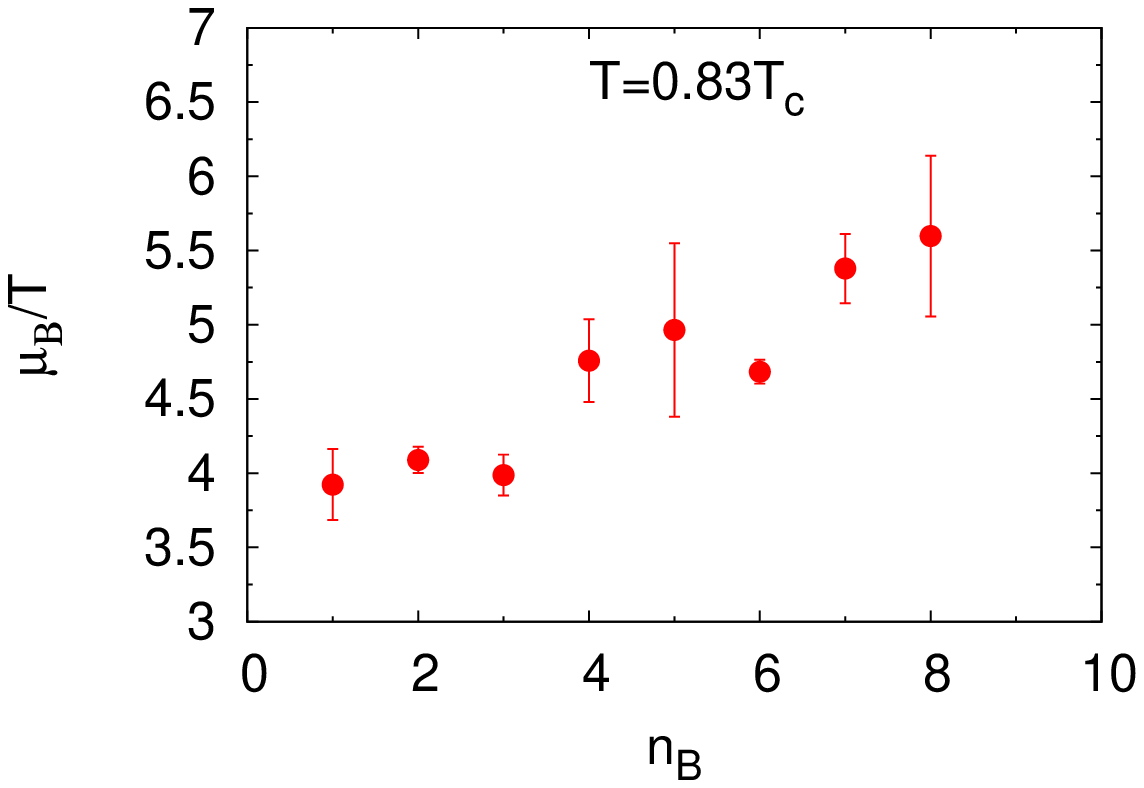} 
\includegraphics[scale=0.39]{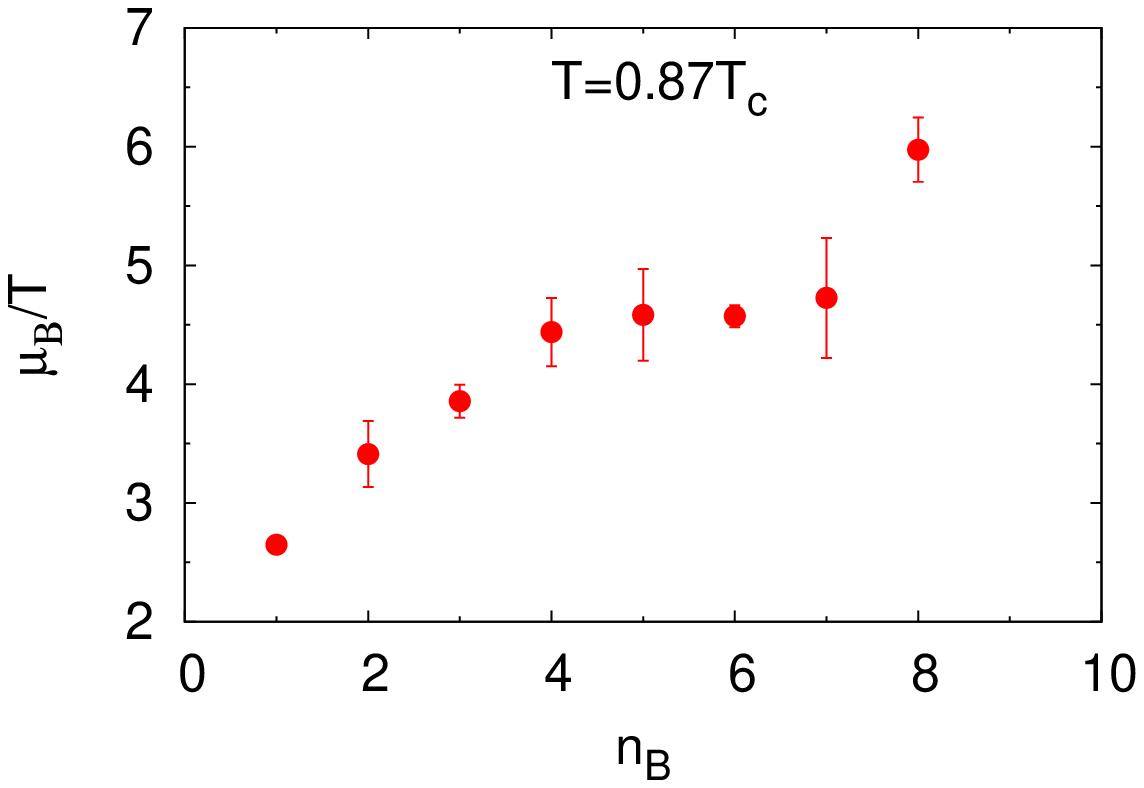} 
\includegraphics[scale=0.39]{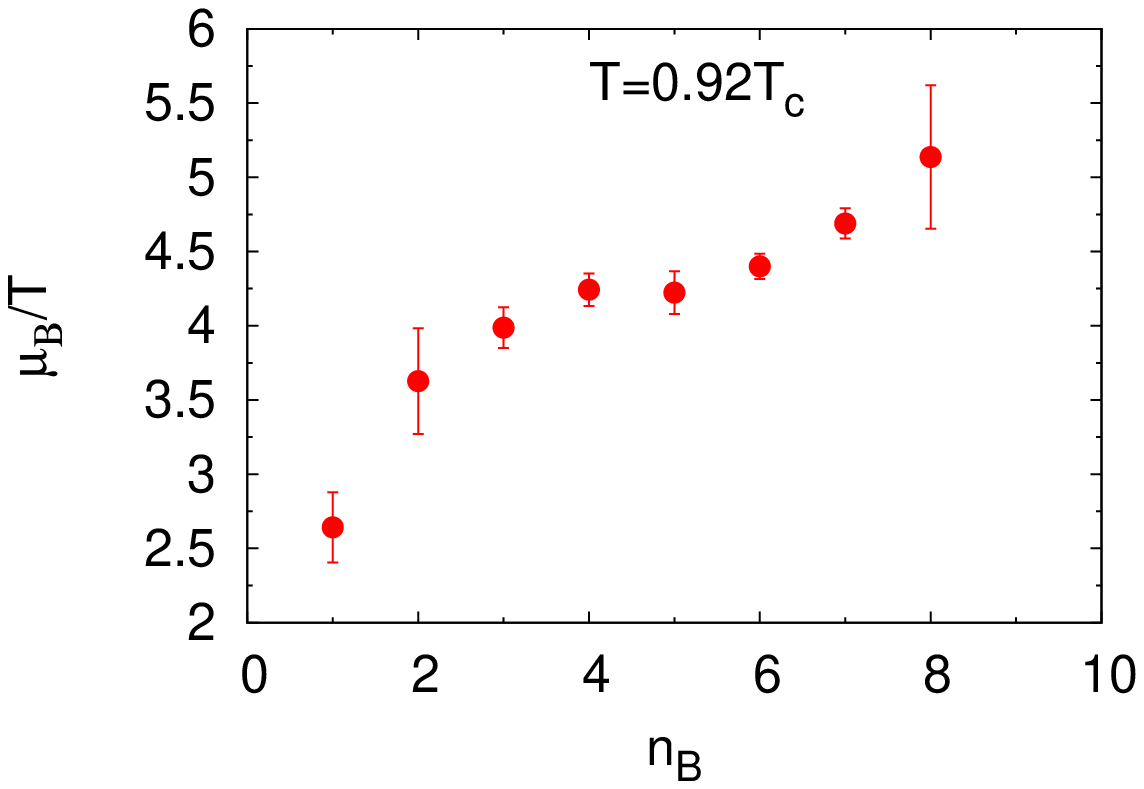}
\caption{Baryon chemical potential for $N_f=2$.}\label{fig:scaning_nf2}
\end{figure}

For the $N_f=2$ case, the phase diagram (see Fig.~\ref{fig:nf2nf4_phase})
is expected to be similar to the conjectured real QCD phase diagram.
For nonzero quark mass, it also shows a crossover behavior at vanishing 
chemical potential, a critical point as well as a first order phase transition line. 
For this system, we scan three temperatures below $\rm{T}_c$: $0.83$, $0.87$ and $0.92{\rm T_c}$. 
The results are presented in Fig.~\ref{fig:scaning_nf2}:
Given our limited statistics of 2000 configurations and the fact that the sign problem starts to set in below $0.83 \rm{T_c}$, we do not observe a clear 
signal for the S-shape structure within the scanned temperature and density range. This may be due to the fact that the transition is milder than the sensitivity of our data, or it may be due to the
fact that the transition occurs at lower temperatures than the ones we studied -- there is at least one study that
claims that the critical point for $N_f=2$ occurs at temperatures below $0.8{\rm T_c}$~\cite{Ejiri:2008xt}.
However, in view of the sign problem we encounter below $0.83 \rm{T_c}$, the claim needs to be scrutinized.

We show the average sign in Eq. (\ref{sign_rew}) in Fig.~\ref{fig:sign_nf24} (right panel). Again, we see that except for the case at the higher $n_B$ and lower temperature,
the average signs for most of the cases are more than three sigmas above zero. 

\subsection{Phase diagram for $N_f=3$}

\begin{figure}[ht!]
\centering
      \includegraphics[scale=0.55]{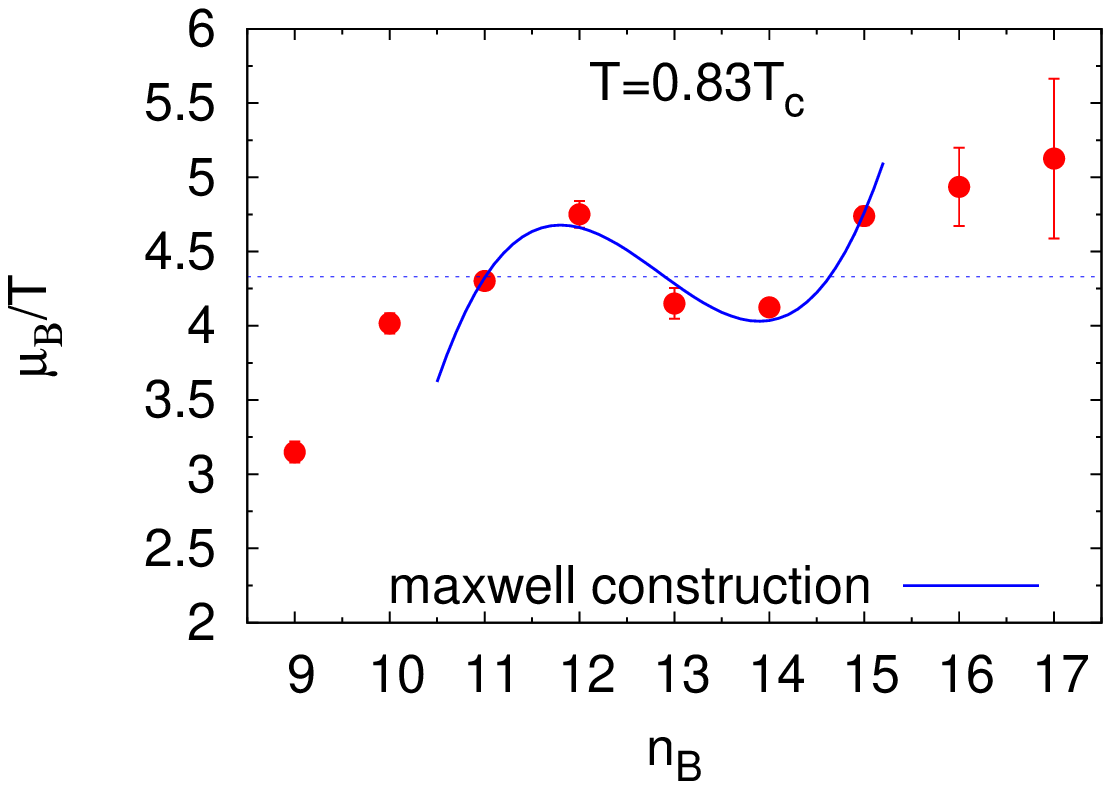}
      \includegraphics[scale=0.55]{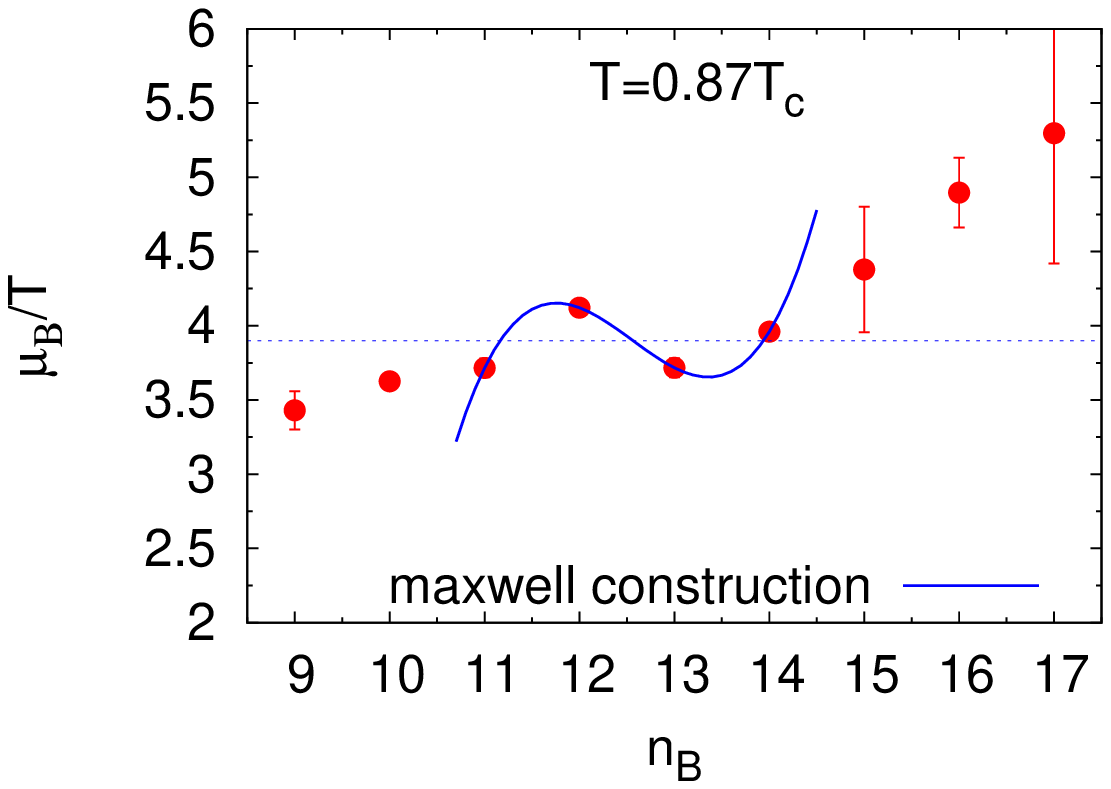}
      \includegraphics[scale=0.55]{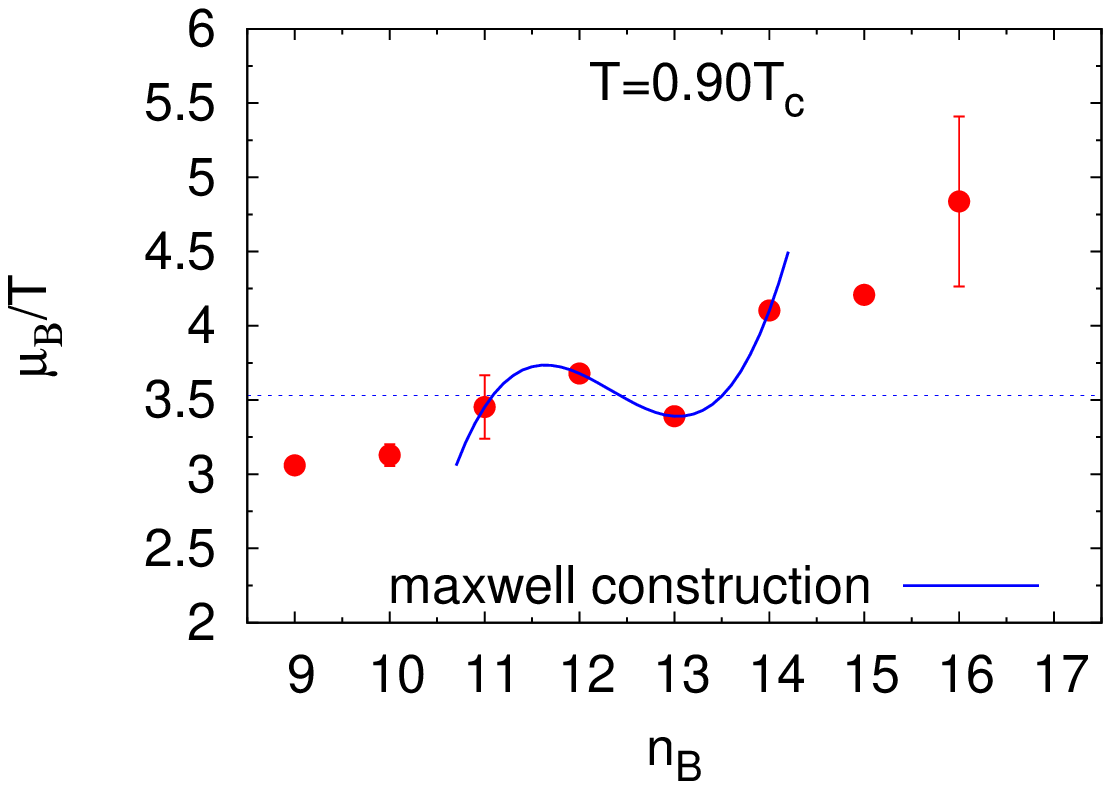}
      \includegraphics[scale=0.55]{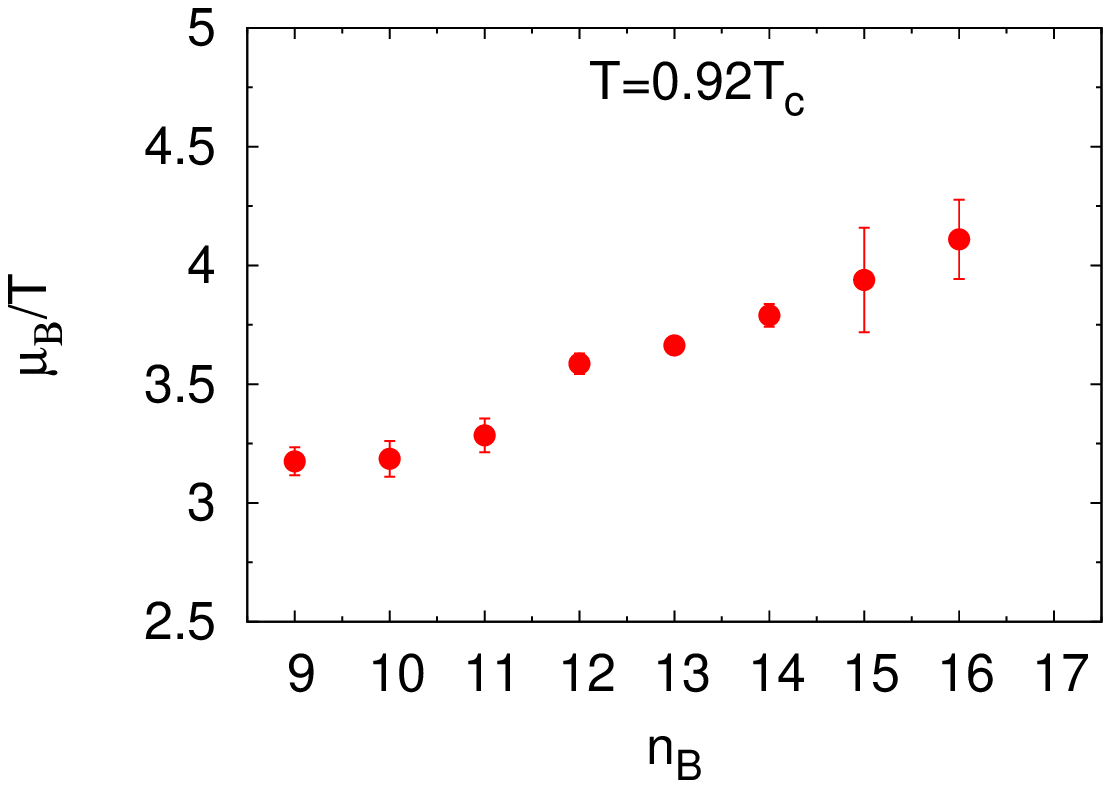}
\caption{$N_f=3$ phase scanning along with Maxwell constructions}\label{fig:scaninng_nf3}
\end{figure}

As we see from the above studies, the canonical approach at finite density is feasible and has been 
checked in the $N_f =4$ case. We would like to extend this study to the real world which contains two 
light quarks and one strange quark. However, simulating light quark masses requires larger volumes 
than the ones used in the present study. As a first attempt, we shall use the quark mass around physical 
strange quark mass as our input and apply the canonical ensemble method to the three degenerated flavor 
case which is expected to have similar features to the real QCD case. 
We are looking for the existence of a first order phase transition and its critical point at this stage, which may 
serve as an indication for the existence and approximate location of such a point in the real world.
Following the methodology for two and four flavors, we fix the temperature at four different values and
scan for the phase diagram by varying $n_B$. The resulting baryon chemical potential is plotted in Fig.~\ref{fig:scaninng_nf3}.

The presence of the S-shape structure at three temperatures below $0.92\rm{T_c}$
is a signal for a first order phase transition. Determination of the coexistence phase 
boundaries is carried out via the Maxwell construction. We plot the results from
the Maxwell constructions in Fig.~\ref{fig:scaninng_nf3}.

\begin{figure}[hbt!]
\centering
\includegraphics[scale=0.55]{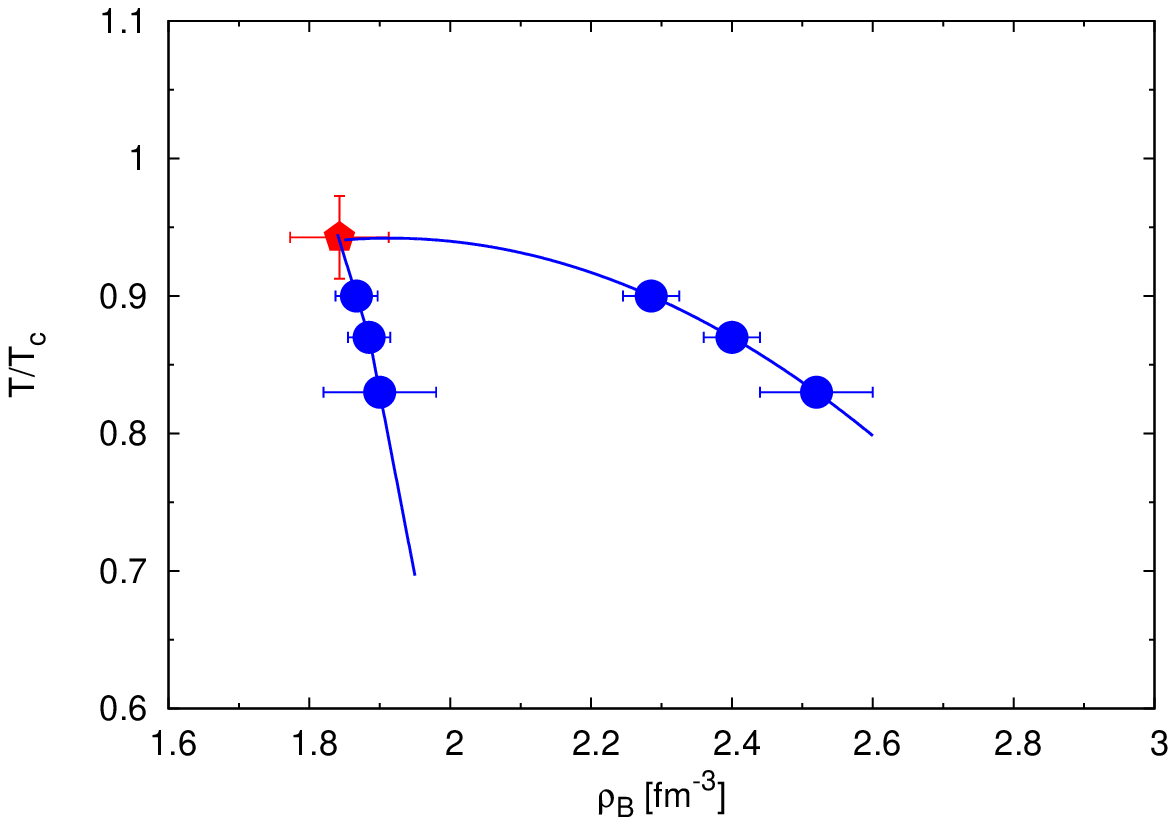}
\includegraphics[scale=0.55]{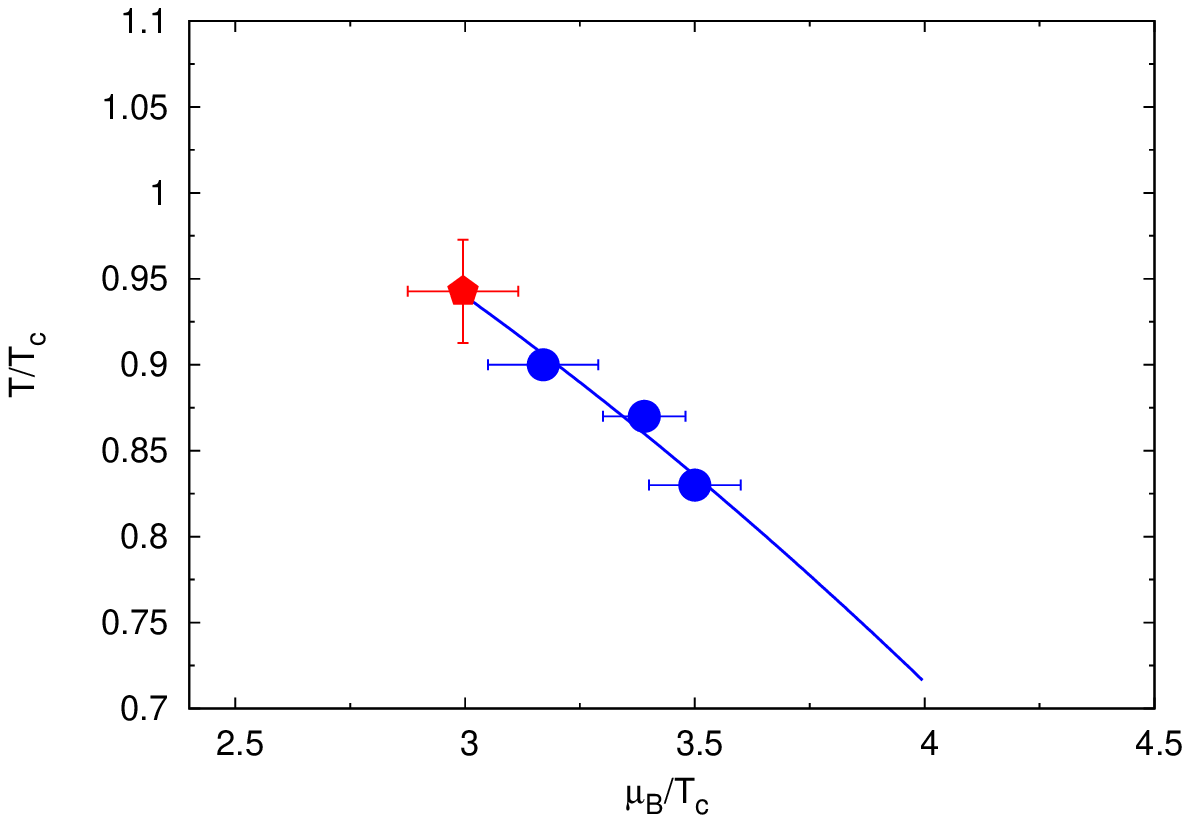}
\caption{Phase boundaries in the temperature vs. density plot for
$N_f=3$ (left). Transition line in the temperature vs. chemical potential plot (right)}\label{fig:bound_nf3}
\end{figure}

In Fig.~\ref{fig:bound_nf3} (left panel) we plot the boundary points as determined by the Maxwell construction.
The phase boundaries intersect at the critical point which is located at $\rm{T}/\rm{T}_c=0.94(3)$ and 
$\rho_B=1.82(7) \rm{fm}^{-3}$. In the right panel of Fig.~\ref{fig:bound_nf3}, we plot the critical chemical 
potential as a function of temperature. On this graph the critical point is located at
$\rm{T_E}=0.94(3)\rm{T_c}$ and $\mu_E=3.01(12) \rm{T_c}$. 

Now that we have determined the phase boundaries in the temperature--density plot,
we could
map out the phase boundary in the temperature--chemical potential plot. The later is the usual phase diagram studied in the grand canonical ensemble. 
It is shown in Fig.~\ref{fig:bound_nf3} (right panel).

\begin{figure}[hbt!]
\centering
      \includegraphics[scale=0.55]{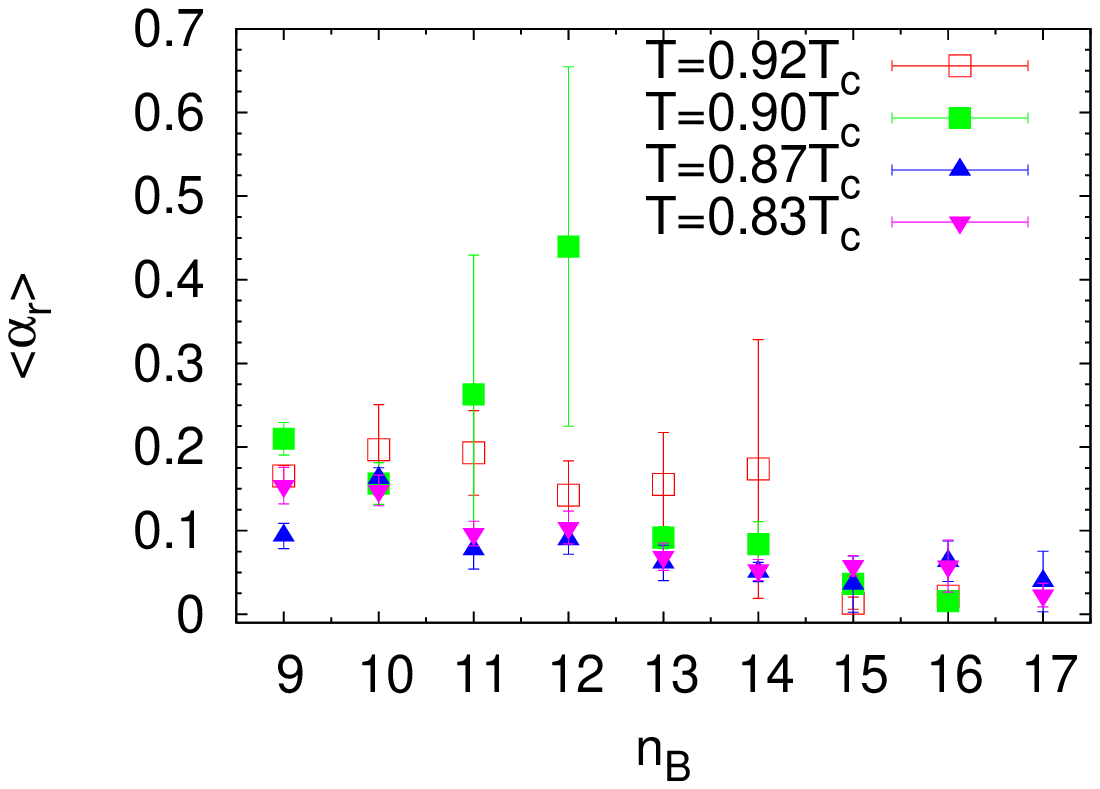}
      \includegraphics[scale=0.55]{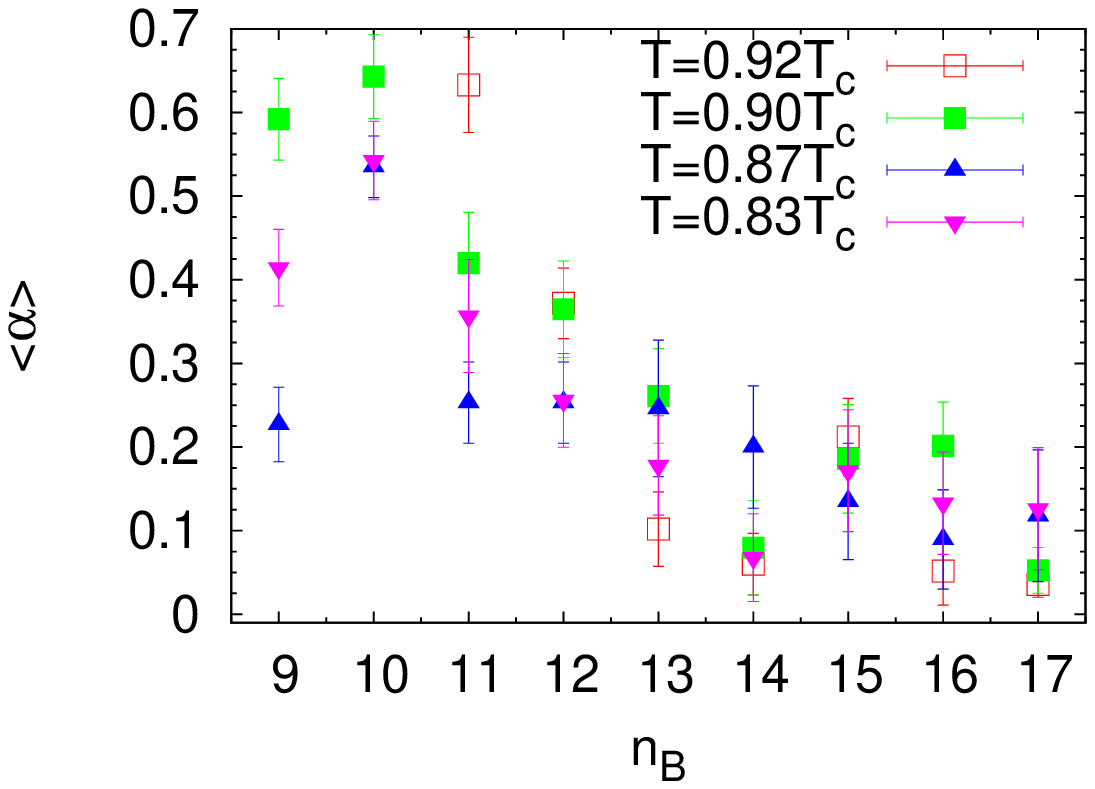}
\caption{Average sign of the $N_f=3$ case for different $n_B$ and $T$ (left: with reweighting; right: without reweighing).}\label{fig:sign_nf3}
\end{figure}

From our current simulations, we have found a first order phase transition and determined the location
of the critical point for the three degenerate flavor case at intermediate quark mass. At the critical point,
the critical temperature is determined as $\rm{T_E}=0.94(3)\rm{T_c}$ and the baryon chemical potential is
 $\mu_E=3.01(12) \rm{T_c}$. Given $\rm{T_E}=0.94(3)\rm{T_c}$, we are supposed to see an S-shaped behavior at $\rm{T}=0.92\rm{T_c}$. However, 
we cannot observe a clear signal of it, it may be due to the fact that this temperature is too close to $\rm{T_E}$.

We show the average sign from reweighting (Eq.~(\ref{sign_rew})) in Fig.~\ref{fig:sign_nf3} (left panel). Again, we see that most of them are more than
3 sigmas above zero. The only exceptions are those points at $n_B > 14$ which are beyond the coexistence phase region. We also plot the
average sign with no reweighting (Eq.~(\ref{sign_def})) in Fig.~\ref{fig:sign_nf3} (right panel). As expected, we find that they are in general larger 
in value than their respective ones with reweighting. 

Before we conclude this section, we would like to discuss a potential
contradiction between our results and those of de Forcrand and 
Philipsen~\cite{deForcrand:2008zi,deForcrand:2008vr,deForcrand:2006hh}.
These authors used 2+1 and 3 flavors staggered fermions and a Taylor expansion in $\mu_q/T$ to study
the curvature of the critical surface at very light quark masses close to $\mu_q=0$ surface. 
After expanding to $(\mu_q/T)^4$, they find that the critical surface bends 
such that the first order region shrinks at higher quark masses. This suggests an exotic scenario 
where there is no critical point at finite chemical potential~\cite{deForcrand:2008zi,deForcrand:2008vr,deForcrand:2006hh}.
This conclusion seems to contradict our results. 
However, since this conclusion is derived from studies at very low chemical potential, it is possible that the critical
surface bends back at larger chemical potentials and the critical point reappears.
In fact, an NJL model study seems to indicate that this scenario is
actually realized~\cite{Chen:2009gv}.

\section{Summary}

In this study, we show that the canonical ensemble can be used to investigate the phase diagram at
finite temperature and nonzero baryon density, at least on small lattices.
In order to scan the QCD phase diagram at relatively large quark number (e.g. $k$ > 20),
we developed the winding number expansion method to calculate the
the projected determinant. It greatly expands
our ability of scanning a broad region of the phase space and 
allows us to reach the coexistence phase.

We presented our results from simulations using two and four degenerate flavors of quarks. 
At finite volume, the first order phase transition at finite density appears as an
S-shaped structure in the plot of chemical potential versus baryon number (baryon density).
To determine the phase boundaries we use the Maxwell construction. For $N_f=4$, 
the first order phase transition is observed at three different
temperatures below the transition temperature at zero chemical potential. The phase boundary
separating the hadron gas phase from the coexistence phase and the
boundary between the plasma phase and the coexistence phase eventually cross at one point.
For the four flavor case, this point should be the first order transition point
at zero chemical potential. We have checked this numerically and found that the intersection 
of the boundaries indeed coincides with the the critical temperature at $\mu=0$ within errors. 
Furthermore, our results are consistent with those of the imaginary chemical potential approach 
in the grand canonical ensemble~\cite{D'Elia:2002gd}. This is a powerful check for our method.
For the two flavor case, we do not see any clear signal of a first order phase transition for 
temperature as low as $0.83 \rm{T_c}$. 

The most interesting study is the three flavor case. We found signals of a first order phase transitions at three
different temperatures and determined the phase boundaries for each temperature.
The critical point located at $\rm{T} = 0.94(3)\rm{T_c}$ and $\mu_B = 3.01(12) \rm{T_c}$ is obtained by an extrapolation. 

We should point out that the present simulations are conducted
on a small volumes, and with relatively large quark masses and a coarse lattice spacing 
($a \sim 0.3$ fm). We plan to run simulations at lower quark masses using nHYP clover fermions.

\acknowledgments
  This proceedings would not been possible without the work from my collaborators of the $\chi$QCD collaboration. I am most grateful to Andrei Alexandru, Keh-Fei Liu and Xiangfei Meng for their significant contributions and enjoyable collaboration. I would also like to thank P. de Forcrand, S. Ejiri, C. Gattringer, F. Karsch and M.-P. Lombardo for their useful discussions and suggestions. This work is partially supported by DOE Grants DE-FG05-84ER40154. 

\end{document}